\documentstyle[twocolumn,eqsecnum,prb,aps,epsf]{revtex}

\def\mysize{8cm}

\catcode`\@=11
\def\rmd{{\rm d}}
\def\rmi{{\rm i}}

\renewcommand{\Re}{{\cal R}\!{\sl e}\,}
\renewcommand{\Im}{{\cal I}\!{\sl m}\,}

\def\JH{J_{\rm H}}
\def\Ham{{\cal H}}

\def\braket#1{\left\langle#1\right\rangle}

\def\simle{\mathrel{\mathpalette\@versim<}}   
\def\simge{\mathrel{\mathpalette\@versim>}}   
\def\@versim#1#2{\lower2.5pt\vbox{\baselineskip0pt \lineskip-.5pt
   \ialign{$\m@th#1\hfil##\hfil$\crcr#2\crcr\sim\crcr}}}

\catcode`\@=12

\begin{document}
\draft

\title{
Scaling relations in charge and spin excitations for La$_{1-x}$Sr$_x$MnO$_3$}
\author{Nobuo Furukawa}

\address{
  Institute for Solid State Physics,  University of Tokyo,
Minato-ku, Tokyo 106-8666, Japan,\\
and\\
  Institute for Theoretical Physics, ETH-H\"onggerberg,
  CH-8093 Zurich, Switzerland
}
\author{Yutaka Moritomo}
\address{
Center for Integrated Research in Science and Engineering (CIRSE),\\
 Nagoya Univ., Nagoya 464-8601 Japan\\
and\\
Precursory Research for Embryonic Scienece and Technology (PRESTO),\\
 Japan Science and Technology Cooperation (JST), 
Chiyoda-ku, Tokyo 102, Japan
}

\author{K. Hirota and Y. Endoh}

\address{
Department of Physics, Tohoku University,\\
  Aramaki Aoba, Sendai 980-77, Japan
}

\date{\today}
\maketitle
\begin{abstract}
Scaling relations in the charge and spin excitations
of La$_{1-x}$Sr$_x$MnO$_3$ 
are studied from both theoretical and experimental points of view.
In the ferromagnetic metal phase, we investigate optical conductivity 
and neutron inelastic scattering, and compare
 with a theoretical calculation 
based on the dynamical mean-field
theory of the double-exchange hamiltonian.
Spin and charge dynamics of La$_{1-x}$Sr$_x$MnO$_3$ exhibit
typical behaviors of half metals.
In these manganite compounds with high Curie temperature,
various behaviors in spin and charge properties are explained by the
double-exchange hamiltonian alone. 
Magnetoresistance of these compounds as well as other
compounds with lower Curie temperature are also discussed.
\end{abstract}



\section{Introduction}

One of the main issues in
colossal magnetoresistance (CMR) manganites ($R$,$A$)MnO$_3$ is
to investigate how to control 
the electronic properties, especially these for transports,
through magnetic field.
In order to  explain the ferromagnetism,
Zener\cite{Zener51} introduced a model hamiltonian 
\begin{equation}
  \Ham = 
  - \sum_{ij,\sigma} t_{ij}
        \left(  c_{i\sigma}^\dagger c_{j\sigma} + h.c. \right)
    -\JH \sum_i \vec\sigma_i \cdot \vec S_i,
    \label{HamSinfty} \label{HamDXM}\label{defDEHam}
\end{equation}
where $t$ is the electron hopping energy and $\JH$ is Hund's coupling
between itinerant electron and localized spins,
and proposed the double-exchange (DE) mechanism.
Anderson and Hasegawa\cite{Anderson55} took the infinite $\JH$ limit
of the model to discuss the ground state spin structure.
However, at finite temperature,
relationship between magnetism and electronic structure
is not fully understood through these calculations.
In the strong Hund's coupling limit  $\JH\gg t$,
mean-field theories which do not  take into account effects of
local spin fluctuations and neglect  changes in
electronic structure and lifetime of conducting quasiparticles
are not sufficient to understand thermodynamic properties
including the Curie temperature $T_{\rm c}$ as well as resistivity.
Therefore, a controlled
method beyond the mean-field theory is
necessary.

From the experimental point of view, recent improvements in 
precise control of the A-site cation substitutions 
revealed a complex phase diagram as a function of substitution,
temperature and magnetic field, with various phases which exhibit 
magnetic, charge, orbital and lattice orderings.\cite{Ramirez97}
Although the DE hamiltonian can describe an antiferromagnet
at half-filling and ferromagnetic metal at doped cases,
it is clear that the model has to be extended in order to
account for charge and orbital orderings.
Theoretically, investigations 
are performed on several microscopic models,
especially emphasis on polaronic models,
to reproduce  varieties of
phases with different behaviors in  resistivity
 and phase transitions.\cite{Bishop97x}
However, due to such a complexity with variety of phases,
there is no satisfactory theory so far to explain
properties in the entire parameter region of substitutions and temperature.

Another point of view is to discuss magnetoresistance (MR)
 of manganites in relation with
other materials such as CrO$_2$ or Tl$_2$Mn$_2$O$_7$.\cite{Khomskii97}
Especially, in various ferromagnetic metals,
the half-metallic behavior due to 
DE interaction  is considered to cause
so-called tunneling magnetoresistance (TMR)
 through the spin valve mechanism.\cite{Hwang96} 
Magnetoresistance of the material with artificially
controlled grain/interface
boundaries are also studied to
realize a low-field MR device through TMR.\cite{Sun96,Steenbeck97,Mathur97}

Thus it is now important to discuss MR in manganites from
the point of view of making distinction between 
generic DE half-metal properties and specific behaviors of these compounds.
In this paper, 
we focus on the bulk nature of the DE systems
in the ferromagnetic metal state.
We study series of compounds which show the metallic ferromagnetism
at the ground state, a canonical examples of which are
La$_{1-x}$Sr$_x$MnO$_3$ with carrier density $x=0.2\sim0.4$,
so-called `high-$T_{\rm c}$' compounds,
and compare with the theoretical results obtained from the DE hamiltonian.
Investigations are through its charge and spin excitation properties.
We present a comprehensive study by putting together theoretical and
experimental results 
which have been described only briefly in previous short papers and letters
by the authors,
and provide a unified picture of the charge and the spin excitation
spectra through the electronic structure of the half-metallic
DE system.
One of our aim is to establish the relevance of the
half-metallic picture for the conduction electrons in these materials,
and investigate the consequence of such a electronic structure
to spin and charge dynamics through the DE mechanism.
We also discuss MR of the low-$T_{\rm c}$ compounds
such as La$_{2/3}$Ca$_{1/3}$MnO$_3$.
In \S2, we show theoretical results for DE Hamiltonian
studied in a nonperturbative manner
using the dynamical mean-field 
approach.\cite{Furukawa94,Furukawa95b,Furukawa95c,Furukawa95d}
In \S3, we show experimental results for spin wave lifetime
with respect to relatively high $T_{\rm c}$ samples
of ((La,Y),Sr)MnO$_3$.\cite{Furukawa98a}
In \S4, we show experimental results for optical spectrum
for (La,Sr)MnO$_3$ as well as Nd substituted
low $T_{\rm c}$ materials.\cite{Moritomo97a}
Section 5 is devoted for discussion and summary.

\section{Theoretical calculation}

\subsection{Model and method}

We consider a DE hamiltonian (\ref{defDEHam}) which is now
also called as ferromagnetic Kondo lattice model.
Since we consider the case where
 the localized spin is in a high-spin state and
the coupling to itinerant electrons is ferromagnetic due to Hund's rule,
the effect of quantum exchange
is speculated to be irrelevant. Especially, near the Curie temperature
$T_{\rm c}$ thermal fluctuations should be dominant.
Thus we make an approximation to replace  the quantum spins operators by
classical rotators ($S=\infty$ limit),
namely we replace the spin operators $\vec{S}_i$ in (\ref{HamDXM}) by 
classical fields with norm $|\vec{S}_i|=1$.

We introduce the dynamical mean-field (DMF) method 
for the above system.
Within this method, the system is treated by a single
site interacting with an electronic bath,m
or the single particle Green's function $\tilde G_0$,
which plays a role of a frequency-dependent Weiss field.\cite{Georges96}
In the limit $S=\infty$ where dynamics of the spin in the imaginary-time
axis does not exist, the single-site problem is exactly solved
albeit the problem must be solved self-consistently with respect to the
external bath. For details, readers are referred 
to ref.~\onlinecite{Furukawa94}.

Advantages of applying this method are: (i) Thermal fluctuations
 of localized spins are taken into account
in a non-perturbative way. Finite temperature calculation in the
thermodynamic limit is possible. (ii) Unlike the case for the Hubbard 
model\cite{Georges96,Pruschke95} the single-site problem is
 solved easily so that no additional approximation is necessary.
The problem is solved in either imaginary frequencies or real frequencies,
so that electronic dynamics are obtained directly.

\subsection{Results}
\subsubsection{Electronic Density of States and Half-Metallic Behaviors}

\begin{figure}
\epsfxsize=\mysize
\hfil\epsfbox{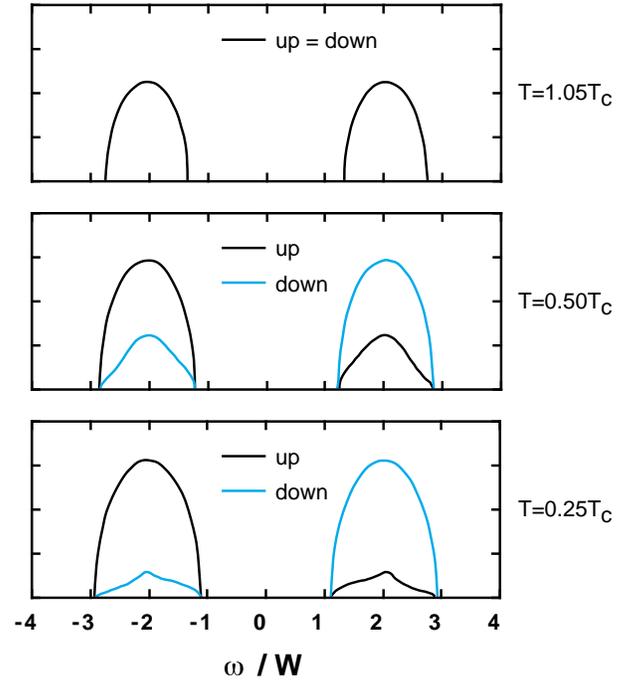}\hfil
\caption{Electron DOS by DMF calculation at
$\JH/W=2$ and $x=0.2$.}
\label{FigDMFDOS}
\end{figure}

In Fig.~\ref{FigDMFDOS} we show the electron DOS $A_\sigma(\omega)$
in DMF on Bethe lattice
at $\JH/W=2$ and $x=0.2$ which gives ferromagnetic ground state.
We set the magnetization to be $+z$ direction.
Results are qualitatively the same as long as the strong 
Hund's coupling region with ferromagnetic ground state is concerned.

In the ground state with ferromagnetic ordering for the localized spins,
Hund's coupling splits the  electron DOS into
upper and lower subbands at $\omega = \pm \JH$,
corresponding to the state parallel and anti-parallel to
the ferromagnetically-ordered local spins.
The width of the each DOS is the bandwidth of the
noninteracting system. Therefore, in the case $2\JH> W$ where
Zeeman splitting is larger than the bandwidth, we have a gap in DOS.
The conduction band is half metallic, namely
the Fermi surface exists only in one of the spin species.

In this system, the origin of the half-metallic behavior 
is due to large Hund's coupling $\JH \gg W$.
This state is in large contrast with itinerant  weak ferromagnet which
has smaller split of DOS  and the Fermi surfaces in both spin species.
In manganites with ferromagnetic metal state,
half-metallic behaviors are observed in
various experiments including photoemission,\cite{Park98a}
tunneling junction\cite{Sun96} and grain boundary TMR 
measurements.\cite{Hwang96}

At finite temperature, the spectra change in such a way that
there exist two peaks at $\omega \sim \pm \JH$ and
its integrated weight transfers by magnetization.
The center of the both peaks split by $\JH$ are roughly unchanged.
Both spin species have finite weight in lower and upper subbands.
For the occupied lower subband,
integrated weight scales as $(1+M^*)/2$ and 
$(1-M^*)/2$ for up (parallel) and down (antiparallel) bands, respectively, 
where $M^* = \langle S_z\rangle / S$ is the magnetization of the local spin.
Due to particle-hole symmetry, the upper subband have the weight
 $(1-M^*)/2$ and $(1+M^*)/2$ for up and down electrons,
respectively.
The width of each sub-bands changes in the way that
at higher temperature the bandwidth is reduced due to
smaller hopping matrix element.

This is easily understood, since the integrated weight of each subbands
gives the spin polarization
of the itinerant electron part, and  
in the strong Hund's coupling
limit it should be proportional to the
spin polarization of the localized spin.
The bandwidth of each subband  becomes narrower
than that for the ground state,
which is understood from the reduction of the hopping matrix 
element.\cite{Anderson55}

Above $T_{\rm c}$, the DOS is still split into two bands,
with the identical weight for both subbands in each spin species.
Within our approach, there is no temperature dependence above $T_{\rm c}$.
This is an artifact of the single-site approximation where
spin correlation of finite length is not taken into account.
However, for strong Hund's coupling,
we speculate that the overall spectrum shape
which is split into two subbands
do not change by finite size cluster corrections.

To investigate the Majority-minority bands more precisely, 
Green's function is calculated as follows.
{}From the solution of the DMF at $\JH/W \gg 1$ on a Lorentzian
DOS system,\cite{Furukawa95c} we have the analytical form for the self-energy
\begin{equation}
  \Sigma_\sigma(\omega) = -\JH - \frac{1-M\sigma}{1+M\sigma}
   (\omega+\mu+\JH + \rmi W)
\end{equation}
at the lower subband $\omega \sim -\JH$ and $\sigma = \pm 1$.
We also see a similar $M$ dependence of $\Sigma$  in semicircular DOS.
Then, the lattice Green's function 
is given in the form
\begin{equation}
  G(k,\omega) = \frac{z_\sigma(M)}
                     { \omega - \zeta_{k,\sigma} + \rmi \Gamma_\sigma}
\end{equation}
where
\begin{eqnarray}
  \zeta_{k\sigma}(M) &=& \frac{1+M\sigma}{2} \varepsilon_k -\JH -\mu,
   \\
  z_\sigma(M) &=& \frac{1 - M\sigma}2,
   \\
  \Gamma_\sigma(M) &= & \frac{1+M\sigma}2 W.
\end{eqnarray}
Since spin-dependent DOS is proportional to
$z_\sigma(M)$, we see the relation $A_\uparrow(\omega) \propto (1+M)/2$ and
$A_\downarrow(\omega)\propto (1-M)/2$.

Let us discuss the low temperature limit $1-M \sim 0$.
For the majority band, we recover
the free-fermion behavior $z=1$, $\zeta_k = \varepsilon_k-\JH-\mu$
and $\Gamma=0$ at the ground state with full polarization $M=1$.
The minority band has different nature.
We see the quasiparticle residue $z(M)\to 0$ and the
 dispersion $\zeta_k = \mbox{const.}$ as $M\to 1$.
This is well understood from Anderson-Hasegawa picture,
that strong $\JH$ 
projects out the low energy minority band and also prohibits
hopping among sites.
What is found here that is not available through
 a simple mean-field treatment of the Anderson-Hasegawa's model
is the lifetime of the minority electrons.
The minority band has large linewidth $\sim W$
due to large amount of scattering with
localized spins in  opposite direction,
so that the propagation of minority electrons are incoherent and 
diffusive. Then we have the width of the minority band $\sim W$
despite real part of the quasi-particle energy becomes 
dispersionless.

\subsubsection{Storner spectrum}

\begin{figure}
\epsfxsize=\mysize
\hfil\epsfbox{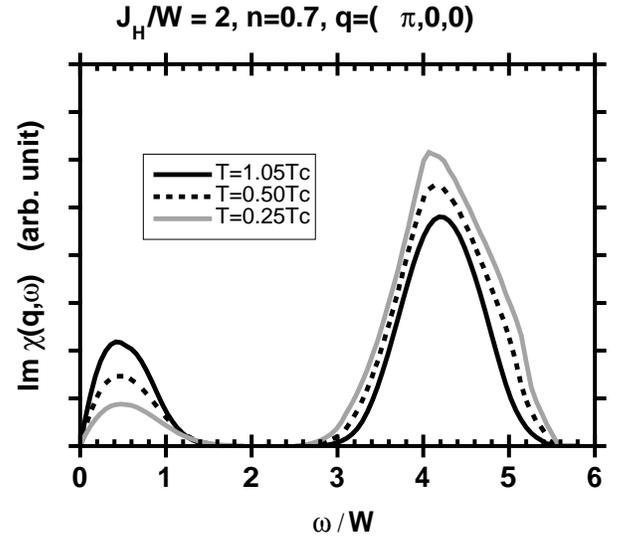}\hfil
\caption{Stoner absorption
${\rm Im}\,\chi(q,\omega)$
 at  $q =
(\pi,0,0)$ for various temperatures.
}
\label{FigImPi}
\end{figure}

Stoner susceptibility is calculated by
\begin{equation}
  \chi(q,z) = \frac1{\beta N}\sum G(k+q,\rmi\omega_n + z ) G(k,\rmi\omega_n).
\end{equation}
Here, correlation effects are taken into account through
the self-energy correction in $G$.
In Fig.~\ref{FigImPi} we show 
${\rm Im}\,\chi(Q,\omega)$
 at the Brillouin zone corner $Q = (\pi,0,0)$ for various temperatures,
at $\JH/W=2$ and  $x=0.3$.
We see two-peak structure at $\omega\sim 0$ and $\omega\sim 2\JH$ which
is explained from the $\JH$-split DOS.
The Stoner absorption is produced from a particle-hole 
pair excitations with spin flip, which produces a peak at low energy
from intra-band processes and another peak at $\omega\sim2\JH$
from interband processes.

\begin{figure}
\epsfxsize=\mysize
\hfil\epsfbox{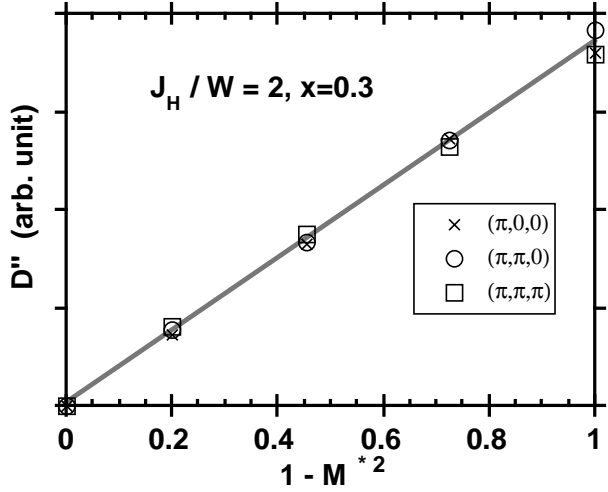}\hfil
\caption{Coefficient for $\omega$-linear part 
$ \partial \Im\chi(Q,\omega)/ \partial \omega |_{\omega\to0}$
as a function of normalized magnetization $M^*$ at 
various wave vectors. Line in the figure is a guide to eyes.
}
\label{FigTheoryScale}
\end{figure}

Let us investigate the low energy part more precisely.
We see that at small $\omega$ we have $\omega$-linear
relation, {\em i.e.} 
$  \Im\chi \propto \omega$ at $\omega \ll W$.
Coefficients for $\omega$-linear part
  decrease by decreasing the temperature. 
In Fig.~\ref{FigTheoryScale} we show 
the coefficient for $\omega$-linear part 
$ D''= \partial \Im\chi(Q,\omega)/ \partial \omega |_{\omega\to0}$
as a function of normalized magnetization $M^*$ at wave vector $Q$.
As a result we find
\begin{equation}
  \Im\chi(Q,\omega) \propto (1-{M^*}^2) \,\omega
   \label{ImChiScale}
\end{equation}
for small values of $\omega$.
The relation (\ref{ImChiScale})
is observed at all values of $q$ with weak $q$ dependence.

This is explained from the fact that the
low energy Stoner absorption is constructed from a process
with combinations of a minority particle and a majority hole carriers.
Since the majority and the minority bands have the spectral weight
proportional to $(1+M)/2$ and $(1-M)/2$, respectively,
the low energy part of the Stoner absorption is proportional to
$1-M^2$. The $\omega$-linear behavior comes from the Fermi distribution
function as is the case for the noninteracting Fermion,
since the
initial and final state is limited in energy $|\epsilon-\mu|<\omega$.
In our case, 
the incoherence of the minority band gives the weak $q$ dependence.
 Thus we have the scaling relation
$\Im\chi \propto (1-{M^*}^2) \omega_q$.
The weak $q$ dependence is in large contrast with the 
conventional weak ferromagnet where minority band is also coherent,
which gives strong $q$ dependence through its band structure.

\subsubsection{Optical spectrum}

Within the framework of the dynamical mean-field theory,
optical conductivity at finite frequencies
is obtained through the Kubo formula
 as
\begin{eqnarray}
  \sigma(\omega) &=& \sigma_0
  \sum_\sigma \int \rmd \omega' \ I_\sigma(\omega',\omega'+\omega)
   \frac{ f(\omega') - f(\omega'+\omega)}{\omega},
   \label{Optcond}
\end{eqnarray}
where
\begin{equation}
   I_\sigma(\omega_1,\omega_2) = \int N_0(\epsilon) \rmd \epsilon \
     W^2 A_\sigma(\epsilon,\omega_1) A_\sigma(\epsilon,\omega_2).
\end{equation}
Here, the spectral weight function $A$ is defined by
\begin{equation}
   A_\sigma(\epsilon,\omega) = 
  -\frac1\pi {\rm Im} G_\sigma(\epsilon,\omega+\rmi\eta) ,
\end{equation}
while $f$ is the Fermi distribution function.
The constant $\sigma_0$ gives the unit of conductivity.

We see  (i) the two-peak structure of $\sigma(\omega)$ at $\omega\sim 0$
and $\omega\sim 2\JH$, and (ii) the transfer of spectral weight 
by temperature. The two-peak structure is due to the splitting of the
DOS at $\pm\JH$. The peak at $\omega\sim0$ corresponds to the
Drude peak which is due to the particle-hole excitation
in the lower subband, while $\omega\sim2\JH$ part is
due to the the interband process between $\JH$-split bands.
At $T>T_{\rm c}$, where DOS is equally split for
up and down spins, the two-peak structure is most pronounce.
In the ferromagnetic phase,
the weight is shifted from the high energy peak to the low energy peak,
corresponding to the DOS structure controlled by the magnetism.
The width of $\sigma(\omega)$ for Drude part becomes narrower
as temperature becomes lower, due to the reduction of the
imaginary part of the self-energy for the majority channel.
The width of the  interband process at $\omega\sim 2\JH$ 
is kept as large as a fraction of $W$,
since this process is transition into the minority channel which
has the incoherent nature with linewidth $\sim W$.

In \S4, we will show the results for $\sigma(\omega)$ in more detail,
in comparison with experimental data,
including the scaling relation for the
transfer of spectral weight as well as the fitting.

\section{Spin excitation and Stoner Absorption}

\subsection{Experimental}

Single crystals of (La$_{1-x-y}$Y$_{y}$)Sr$_{x}$MnO$_{3}$ (typically
$6\phi \times 80$~mm) were grown by the floating-zone method.  The end part
(30--40~mm) were used for neutron measurements.  In order to
study the effects of electronic band-width, three samples were chosen; $y=0.00,
0.05$ and 0.10 with the Sr concentration fixed at
$x=0.20$.  The tolerance factor $t$, which is defined as
$t=(r_{A}+r_{O})/\protect\sqrt{2}(r_{B}+r_{O})$ for ABO$_{3}$
perovskites and scales with the band width, decreases with
doping Y; 0.908, 0.906 and 0.903 for $y=0.00, 0.05$ and 0.10, respectively. 
These values are, however, still closer to unity than
La$_{0.8}$Ca$_{0.2}$MnO$_{3}$ (0.894) and Pr$_{0.8}$Ca$_{0.2}$MnO$_{3}$
(0.877). 
 Neutron-scattering measurements were carried out using the triple-axis
spectrometer TOPAN at the JRR-3M reactor in Japan Atomic Energy Research
Institute. Typical condition employed was the fixed final energy at 14.7~meV
with horizontal collimation of Blank-30$'$-Sample-60$'$-Blank.  The
$(002)$ reflection of pyrolytic graphite (PG) is used to monochromatize and
analyze neutrons.  A PG filter was used to reduce higher-order contaminations
in the incident beam.  The Curie temperature $T_{C}$ and its Gaussian
distribution $\Delta T_{C}$ was determined using the temperature dependence of
the $(100)_{Cubic}$ peak intensity; $T_{C}$ ($\Delta T_{C}$) is 306(1.1),
281(1.4) and 271(10)~K for $y=0.00, 0.05$ and 0.10, respectively.

\begin{figure}
\epsfxsize=\mysize
\hfil\epsfbox{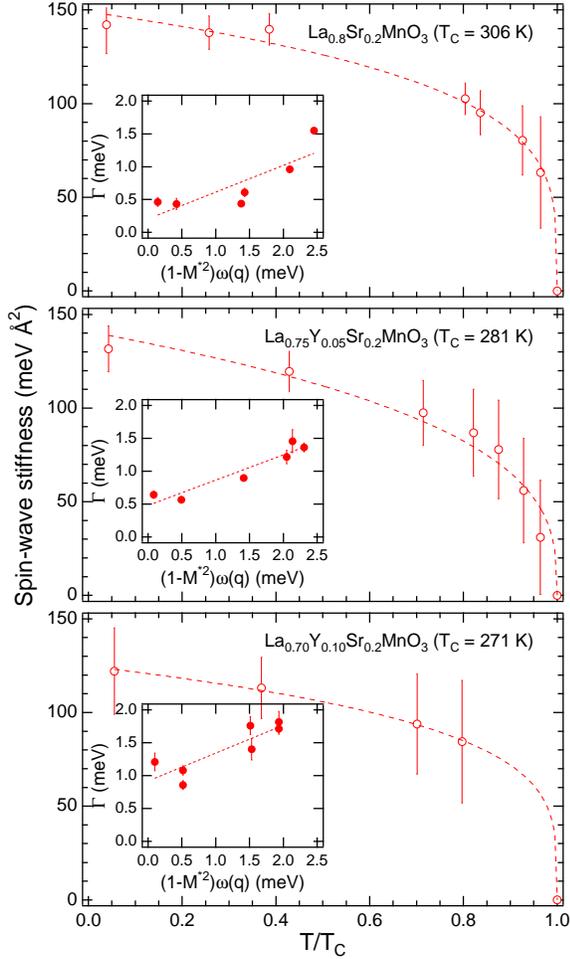}\hfil
 \caption{Experimental results for the 
 temperature dependence of the spin-wave stiffness $D$.  Error bars
 indicate fitting errors.  Dashed lines are guides to the eye.  Inset shows a
 linear relation between $\Gamma(q,T)$ and $(1-M^{*2})\omega(q,T)$, where
 $M^{*}=M(T)/M(0)$, at $q=(1.1, 1.1, 0)$ in rlu.}
 \label{FigStiffness}
\end{figure}

The spin-wave dispersion curves for studied
(La$_{1-x-y}$Y$_{y}$)Sr$_{x}$MnO$_{3}$ samples show 
an isotropic behavior in the
measured ($q$, $\omega$) range,
$q<0.40$~\AA$^{-1}$ and $\hbar\omega<20$~meV, and follow well with
$\hbar\omega(q)=Dq^{2}+E_{0}$, typical of ferromagnetic spin-wave dispersion in
low $q$ and low $\omega$ range.  In order to study the temperature
dependence of spin-wave stiffness $D$, we have performed constant-$Q$
scans at (1.1\ 1.1\ 0) where well-defined peak profiles are obtained in a wide
temperature range. Each peak profile was fitted with the spin-wave
scattering cross section including the finite life-time $h/\Gamma_q$ convoluted
with a proper instrumental resolution. Figure \ref{FigStiffness} shows thus
obtained temperature dependence of $D$.  Error bars
indicate fitting errors.  Dashed lines are guides to the eye.
Softening of spin wave dispersion $\omega(q,T)$ is observed
 as temperature approaches $T_{\rm c}$.

\subsection{Scaling relation}

Spin wave dispersion relation throughout the
brillouine zone as well as the temperature dependence of the magnon linewidth
is first reported by Perring {\em et al.}
for La$_{0.7}$Pb$_{0.3}$MnO$_3$.\cite{Perring96}
The cosine-band like dispersion as well as its slight softening
at the zone boundary is reproduced by the spin wave 
expansion of the double-exchange model,\cite{Furukawa96}
and  a qualitative explanation for the linewidth is
through the Stoner absorption mechanism is also discussed there.
In this paper, we describe  the temperature dependence of the  linewidth
more precisely through a scaling relation between
the magnetization and the magnon linewidth.

Within the linear spin wave theory of the double-exchange model
at $\JH \gg W$,\cite{Furukawa96}
 the dispersion relation $\omega_q$
as well as its linewidth $\Gamma_q$ is obtained by
\begin{eqnarray}
  \omega_q &=& \Re \Pi (q,\omega_q),\\
  \Gamma_q &=& \Im \Pi (q,\omega_q),
\end{eqnarray}
where the magnon self-energy $\Pi$ is calculated by
\begin{equation}
  \Pi(q,\omega) = (\JH^2/St) \chi(q,\omega),
\end{equation}
and $\chi(q,\omega)$ is the Stoner susceptibility.
Note that in the large Hund's coupling region
 the magnon bandwidth $J_{\rm magnon}$ is scaled by $W$ 
as is the case for $T_{\rm c}$.
Namely, in the region $\JH/W \gg 1$ the leading order of
 $J_{\rm magnon} / W$ is $1/S$ and does not contain $\JH$
 due to the nature of Hund's coupling as a projection.
Indeed, the result is identical in the limit $\JH\to\infty$
treated by Kubo and Ohata.\cite{Kubo72}
This justifies the linear spin wave theory at large $\JH$ region,
 at least asymptotically in the large $1/S$ limit.

Higher order terms in the asymptotic $1/S$ expansion
give vertex corrections. However, vertex corrections
may be neglected by restricting ourselves
in the low temperature region where magnon density is low,
and in the sufficiently large doping concentration region
where electron kinetic energy is larger than the magnon energy.
Accuracy of the linear spin wave calculation in the sufficiently
doped region is shown by comparison with $S=1/2$ model,\cite{Kaplan97}
which at the same time shows the softening of the zone boundary magnon
dispersion when electron kinetic energy is small at low electron/hole
doping.

Let us now assume that the major origin of the linewidth broadening
is due to Stoner absorption mechanism, which implies that
we neglect magnon-magnon interactions as well as other mechanisms of
 extrinsic origins.
{}From eq.~(\ref{ImChiScale}) we obtain
\begin{equation}
  \Gamma_q = \alpha_q (1-{M^*}^2) \, \omega_q
	\label{Gamma-scale}
\end{equation}
where $\alpha_q$ is a dimensionless function of $q$.
In other words, the definition of  the spin stiffness is extended
into a complex number
$D= D' + {\rm i} D''$ and the dispersion as well as its linewidth
at small $q$ region is expressed as
$\omega_q = D' q^2$ and $\Gamma_q = D'' q^2$.

As for the experiment,
 we show the inverse lifetime
$\Gamma(q,T)$ versus $(1-{M^*}^2)\, \omega(q,T)$
at the inset of Fig.~\ref{FigStiffness}.
{}From the figure we see that inverse lifetime is fitted in the form
\begin{equation}
  \Gamma(q,T) = \Gamma_0(q) + \alpha_q (1-{M^*}^2) \,\omega(q,T).
\end{equation}
Temperature dependence  is scaled in a way consistent with theoretical
result for DE systems. 
$\Gamma_0$ is the $T$-independent part
of the spin wave lifetime, {\em i.e.} $\Gamma(q,T\to0) = \Gamma_0(q)$,
 which increases  systematically by increasing the ratio of Y atoms.

From the data we consider that the temperature dependent part of the
magnon damping mechanism is mainly due to Stoner absorption which
does not show large change by Y doping. 
Large change by Y doping is seen in the temperature independent 
part $\Gamma_0$.
We speculate that the origin of
 $\Gamma_0$ is mainly extrinsic effects
such as disorder and inhomogeneity of the sample.
Substitution of La by Y creates the decrease of $\braket{r_{\rm A}}$
which causes a reduction of the average electron hopping, as well as
the local Mn-O-Mn bond distortion. In both cases,
 they contribute to
a charge inhomogeneity through
 charge segregation\cite{Yunoki98,Kagan98x,Nagaev97}
or localization. Since spin stiffness is a function of charge concentration,
especially $D\propto x$ in the region of interest here,\cite{Hirota96}
charge inhomogeneity creates distribution of spin stiffness in
a mesoscopic scale.
Inhomogeneities in magnetism as well as charge
distributions are also reported in (La,Ca)MnO$_3$ by various
measurements 
including $\mu$SR,\cite{Heffner96} Raman \cite{Yoon98}
and photoemission.\cite{Booth98}
Broadening of the spin wave linewidth at the zone boundary observed
at lower $T_{\rm c}$ compounds\cite{Hwang98} might also indicate
the inhomogeneities in mesoscopic length scale.

\section{Charge excitation between the $\JH$-split bands}

\subsection{Experimental}
Thin films of $R_{0.6}$Sr$_{0.4}$MnO$_3$ ($R$=La, La$_{0.5}$Nd$_{0.5}$,
Nd$_{0.5}$Sm$_{0.5}$ and Nd$_{0.25}$Sm$_{0.75}$) with thickness of
$\sim$100 nm were fabricated using a vacuum pulsed laser deposition (PLD)
apparatus.
Details of film synthesis are described in Ref.~\onlinecite{Moritomo97a}.
The compound was deposited on to a MgO (100) substrate, since the substrate
is transparent in a wide energy region of 0.1 - 5.0 eV and is suitable for
the absorption measurements.
An absorption coefficient $\alpha$ was determined from transmission spectra
using the standard formula neglecting a multi-reflection effect, since the
optical density of our films is larger than 0.7 in the spectral region
investigated.
X-ray diffraction measurements revealed that the obtained films were
(110)-oriented in the pseudo-cubic setting.
Thicknesses of the films were measured by a scanning electron microscope (SEM). 

Temperature dependence of magnetization $M$ was measured under a field of
0.5 T after cooling down to 5 K in zero field (ZFC), using a
superconducting quantum interference device (SQUID) magnetometer.
$T_{\rm C}$ was determined from the inflection point of the $M-T$ curve.
With decreasing $r_A$, $T_{\rm C}$ decreases from $\approx$325 K for
$R$=La, $\approx$315 K for Nd$_{0.5}$Sr$_{0.5}$, $\approx$175 K for
Nd$_{0.5}$Sm$_{0.5}$ to $\approx$90 K for Nd$_{0.25}$Sm$_{0.75}$.
The suppression of $T_{\rm C}$ is originated in reduction of the Mn-O-Mn
bond angle, and hence the transfer integral $t$ ({\it chemical pressure}
effect\cite{Hwang95,Moritomo97}).
In other words, we can control the $W$-value with changing the averaged
ionic radius $r_{\rm R}$ of the trivalent rare-earth atom $R^{3+}$.

\subsection{Temperature-dependent absorption spectra}
\begin{figure}
\epsfxsize=\mysize
\hfil\epsfbox{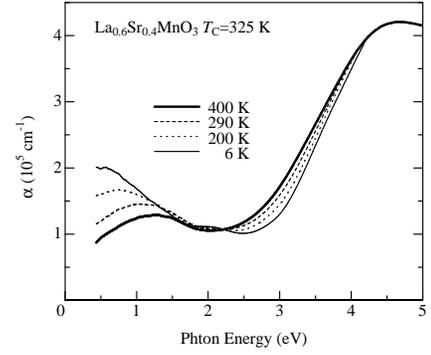}\hfil
\caption{Temperature dependence of absorption spectrum $\alpha(\omega)$ for
La$_{0.4}$Sr$_{0.6}$MnO$_3$ film. The hatched area ($\alpha_{\rm CT}$) is
due to the charge-transfer transition.}
\label{sigma}
\end{figure}

\begin{figure}
\epsfxsize=\mysize
\hfil\epsfbox{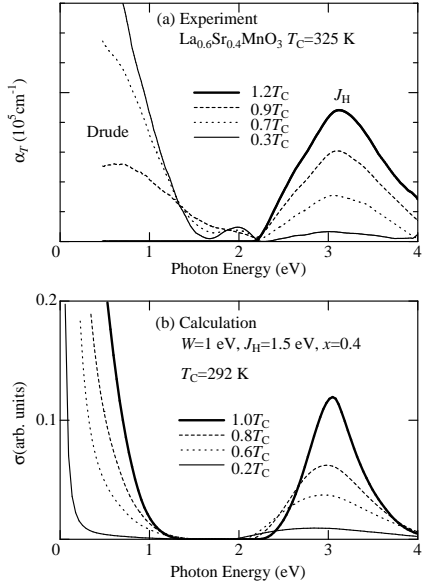}\hfil
\caption{(a) Temperature-dependent component of the absorption spectrum
$\alpha_T$($\omega$)
 for
La$_{0.6}$Sr$_{0.4}$MnO$_{3}$. The bands denoted by $\JH$ is due to interband
transition between the $\JH$-split bands. (b) Calculated results based on the
double-exchange model with DMF and $S$=$\infty$ approximation. The
used parameters are $W$=1 eV, $\JH$=1.5 eV and $x$=0.4.}
\label{sigmat}
\end{figure}

Figure~\ref{sigma} shows variation of the absorption spectrum $\alpha
(\omega)$ (=$(4\pi/cn(\omega))\cdot \sigma(\omega))$ in the temperature
range of 6 - 400 K for La$_{0.6}$Sr$_{0.4}$MnO$_3$ with maximal-$W$
($T_{\rm C}\approx$325 K).
Here, refractive index $n(\omega)$ is almost constant, and hence
$\alpha(\omega)\propto\sigma(\omega)$, above $\sim$0.5 eV.
With decreasing temperature, a large transfer of spectral weight from
$\sim$3 eV to $\sim$0 eV is observed.
This has been interpreted as the spectral weight transfer from the
interband transition between the $\JH$-split bands ($\JH$-gap excitation) to
the intraband transition within the lower band\cite{Moritomo97a}.
To analyze the spectral behavior more in detail, we extract the temperature
dependent component $\alpha_T(\omega)$:
\begin{equation}
       \alpha_T(\omega) = \alpha(\omega) - \alpha_{\rm CT}(\omega).
\end{equation}
The temperature-independent component $\alpha_{\rm CT}(\omega)$ (hatched
region in Fig.\ref{sigma}) is ascribed to the charge-transfer (CT) type
transitions from ligand $p$ states to Mn 3$d$ levels\cite{Moritomo97a}.
Figure~\ref{sigmat}(a) shows the $\alpha_T(\omega)$ spectra for
La$_{0.6}$Sr$_{0.4}$MnO$_3$.
The absorption band located at $\sim$3 eV, which corresponds to the $\JH$-gap
excitation, significantly decreases in intensity with decreasing
temperature.
This is because the density of state (DOS) of the up- (down-) spin states
in the upper (lower) $\JH$-split band decreases with $t_{2g}$-spin
polarization.
The lost spectral weight is transferred into the lower-lying Drude
component ($\leq$1 eV).
Similar behaviors are observed in the spectra for the small-$W$ compounds,
{\it e.g.}, $R$=Nd$_{0.5}$Sm$_{0.5}$ and Nd$_{0.25}$Sm$_{0.75}$, though a
small polaron band becomes discernible at $\sim$1.5 eV.\cite{Machida97x}

The temperature-dependent behavior of the $\JH$-gap excitation is well
reproduced by the double-exchange model that include only $\JH$- and
$t$-terms.
Figure~\ref{sigmat}(b) represents the calculated optical conductivity
$\sigma(\omega)$ with $n$=$\infty$ and $S$=$\infty$ approximation.
The two parameters, i.e., $W$(=1 eV) and $\JH$(=1.5 eV), are estimated from
the $\JH$-gap band in Fig.\ref{sigmat}(b).
As seen in Fig.\ref{sigmat}, the calculated results well reproduce not only
the overall profile but also the temperature dependence of the $\JH$-gap band
around $\sim$3 eV.
Note that the estimate for $T_{\rm C}$ in the present model (=292 K)
 is very close to the
experimentally derived value (=325 K).

In the low energy Drude region ($\leq$1.5 eV), however, the
temperature-dependent behavior of $\alpha_{\rm T}(\omega)$
(Fig.\ref{sigmat}(a)) is qualitative different from the calculation
(Fig.\ref{sigmat}(b)).
The Drude component monotonously increases with decreasing temperature, but
does not show any sharpening behavior like the calculation.
The discrepancy suggests that additional factors, such as dynamical
Jahn-Teller effect\cite{Millis96l,Millis96b} or degeneracy of the
$e_g$-orbitals\cite{Shiba97,deBrito98},
might be necessary to reproduce the low energy charge
excitation.

\subsection{Scaling relation}

\begin{figure}
\epsfxsize=\mysize
\hfil\epsfbox{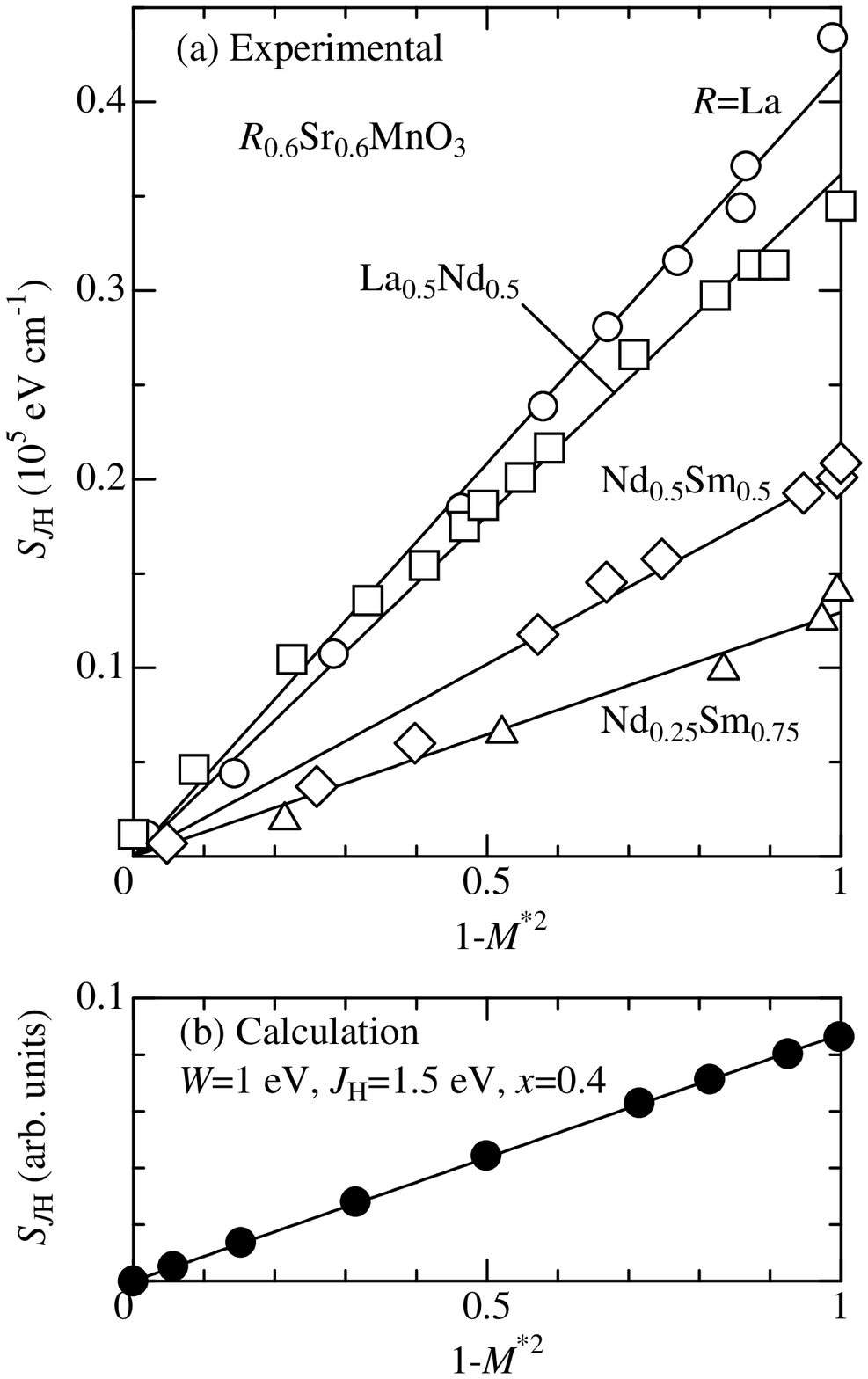}\hfil


\caption{(a) Spectral weight $S_{\JH}$
 of the $\JH$-gap transition against $1-{M^*}^2$,
where $M^*=M(T)/M(0)$ is the magnetization normalized by its saturation value.
(b) Calculated results based on the double-exchange model
with DMF and $S=\infty$ approximation ($W$=1 eV, $\JH$=1.5 eV and
$x$=0.4).}
\label{sj}
\end{figure}

In doped manganites, DOS of the $e_g$-electrons is governed by confirmation
of the local $t_{2g}$-spin due to the strong on-site change interaction
$\JH$.
Therefore, the apparent temperature dependence of $\alpha_{\rm T}(\omega)$
spectrum (Fig.\ref{sigmat}(a)) should be ascribed to the induced
magnetization $M^*$.
In Fig.\ref{sj}(a), we plot the spectral weight of the $\JH$-gap band,
\begin{equation}
  S_{\JH}\equiv\int_{\rm 2.2eV}^{\rm 4.0eV}\alpha_T(\omega )d\omega,
\end{equation}
for $R_{0.6}$Sr$_{0.4}$MnO$_3$ as a function of $1-M^*{}^2$.
The values of $M^*=M(T)/M(0)$ 
were estimated under a field of 0.5 T to avoid the
complexity arising from the nonremanent behavior of the present system.
Change of the $S_{\JH}$-value with temperature scales well 
with $1-M^*{}^2$.
As shown in the Fig.\ref{sj}(b), above scaling relation is reproduced by
calculation based on the double exchange model,
\begin{equation}
  S_{\rm theory} = \int_{\omega\sim 2\JH} \sigma(\omega) {\rm d}\omega,
   \label{Sj-theory}
\end{equation}
where integration is performed at the high energy peak region
around $\omega\sim 2\JH$.

These scaling relation 
(\ref{Sj-scale}) and (\ref{Sj-theory}) are explained as follows.
For the charge excitations,
interband optical absorption is constructed by
spin conserving process of scattering lower band electrons
into upper bands. The initial and final state weight is
again $(1+M^*)/2$ and $(1-M^*)/2$, respectively,
and the optical absorption is proportional to $(1-{M^*}^2)$.
Phenomenologically, this is because the DOS of the up- (down-) spin states
in the lower (upper) $\JH$-split band increases as $\propto(1+M^*)$,
while the up- (down-) spin states in the upper (lower) band decreases as
$\propto(1-M^*)$.

Finally, let us comment on the strength of the $\JH$-gap band.
The spectral weight $S_{\JH}$ is expressed as:
\begin{equation}
  S_{\JH}=(2\pi^2e^2N_0/cm_0\overline{n})\cdot f_0\cdot(1-{M^*}^2),
  \label{Sj-scale}
\end{equation}
where averaged refractive index $\overline{n}$ is $\approx 2$ around the
$\JH$-gap transition.
$f_0$ is the oscillator strength for the $d-d$ transition per one
$e_g$-electron.
As seen in Fig.~\ref{sj}, the magnitude of $S_{\JH}$ at $M^*\approx$0
(high temperature limit) is strongly material dependent, indicating that
the $f_0$-value increases from $\approx$0.04 for $R$=Nd$_{0.5}$Sm$_{0.5}$
to $\approx$0.08 for La.
Remember that $f_0$ is proportional to square of the transition matrix
element between the neighboring $e_g$-orbitals, which is nearly
proportional to $t$, and hence $W$.
Recently, Machida {\it et al.}\cite{Machida97x} have evaluated the $W$-values
for $R_{0.6}$Sr$_{0.4}$MnO$_3$ from $\alpha(\omega)$ spectra: $W$ increases
from $\approx$0.6 eV for $R$=Nd$_{0.5}$Sm$_{0.5}$ to $\approx$0.8 eV for
$R$=La.
Such a material-dependent $W$ qualitatively explain the observed
enhancement of the $f_0$-value.

\section{Discussion}

The general role of the DE mechanism to the electronic structure is
understood through the half-metallic DOS at the ground state
and the shift of the spectral weight by the magnetic fluctuations.
As a consequence of the DOS structure, spin and charge dynamics
are influenced through a particle-hole channel.
In experiments,  optical conductivity $\sigma(\omega)$
and spin wave linewidth $\Gamma_q$ show scaling relations
in agreement with the theory. 

Let us discuss the two different 
energy scale of the DE system and CMR materials which
arises from DOS structure controlled by magnetism.
The electronic structure changes its spectrum by magnetization
in the scale of $2\JH \sim 3 {\rm eV}$ if both of the
$\JH$-split band are concerned. The scale of the change in the spectrum
for a case when only the  lower sub-band is playing a role is
$W\sim 1{\rm eV}$. However, the energy scale concerning
the magnetic properties has a smaller energy scale determined
by the DE interaction, $J_{\rm DE} \sim T_{\rm c} \sim 10^{-2}{\rm eV}$.
Thus a small change in magnetization causes a large change in
electronic structures, {\em e.g.} a change of temperature
in the scale of 300K causes the change in optical spectrum at 3eV.
Such a huge change in DOS spectrum  controlled by
a small amount of magnetic field is considered to
cause sensitive change of transport properties
as well as other thermodynamical properties.\cite{Asamitsu95}

Similar phenomena happens for a correlated electron system
in the vicinity of  the Mott transition.\cite{Rozenberg95}
 This case, the coherent temperature
becomes small compared to the bandwidth, $T_{\rm coh}\ll W$,
so that the change in the scale of $T_{\rm coh}$ produce
a large change in the electronic structures.
For the DE systems, however,  
the small energy scale $J_{\rm DE}$ and the large energy
scale $W$ as well as $\JH$ coexists robustly,
and such phenomena is commonly observed throughout the doping concentration.

In order to compare the theory with CMR compounds,
we need to consider that
their behavior changes by A-site substitutions.
Let us concentrate to the region $x\sim 1/3$ where
it is far from antiferromagnetic insulating phase at $x\sim 0$
and the region with charge and orbital ordering at $x\sim 0.5$.
The nature of the compound is roughly classified by the
average radius of the A-site ions
 $\langle r_{\rm A}\rangle$.\cite{Hwang95,Moritomo97,Radaelli97}
Compounds with larger $\langle r_{\rm A}\rangle$ 
have to larger tolerance factor and larger Mn-O-Mn bond angle,
which in general increase the effective electron hopping
and hence $T_{\rm c}$.
However, we should note that
 it is still controversial  whether the phase diagram is 
controlled mostly by the bandwidth alone, {\em e.g.}
ionic size variation $\sigma( r_{\rm A})$
also plays some role to change $T_{\rm c}$,\cite{Rodriguez-Martinez96}
and also the decrease of $T_{\rm c}$ at lower $T_{\rm c}$ region
is much larger compared to the estimate from 
the change in $\langle r_{\rm A}\rangle$.

A canonical example for the high $T_{\rm c}$ compounds is
(La,Sr)MnO$_3$ which is resistivity wise a good metal below $T_{\rm c}$,
and incoherent metal with $\rho(T)$ near Mott's limit.
The most well investigated compound (La,Ca)MnO$_3$, where Ca
substitution creates smaller $\braket{r_{\rm A}}$ and larger 
$\sigma(r_{\rm A})$, has lower $T_{\rm c}$,
and shows metal to insulator transition at around $T_{\rm c}$.
Compounds with smaller $\braket{r_{\rm A}}$ such as (Pr,Ca)MnO$_3$
do not undergo ferromagnetic transition and stay insulating.\cite{Ramirez97}

As long as  the wide-banded compound with high doping are
concerned, {\em e.g.} (La,Sr)MnO$_3$ at $x = 0.2 \sim 0.4$,
the DE Hamiltonian accounts for several 
experimental data concerning the ferromagnetism and 
the transport.
Curie temperature is estimated by various
 methods\cite{Furukawa95b,Yunoki98,Roder97}
and is consistent with experiments in magnitude as well as
the doping dependence.
The resistivity is calculated by
the dynamical mean-field approach\cite{Furukawa94,Furukawa95d}
and the result is consistent with the experimental data 
for resistivity\cite{Tokura94,Urushibara95} which show metallic behavior
$\rmd \rho / \rmd T >0$ at $T>T_{\rm c}$
as well as the absolute value $\rho(T) = 2 \sim 4 {\rm m}\Omega{\rm cm}$
which is not so large compared to the Mott limit.
Universal behavior of the magnetoresistance in the form
$  -\Delta\rho/\rho_0\propto   {M^*}^2$
is also consistent between theory and experiment.

Together with the scaling relations described in this paper,
DE Hamiltonian alone explains various experimental data in
manganites with high $T_{\rm c}$.
The  mechanism of MR in the single-crystal of these
 high $T_{\rm c}$ compounds are understood from double-exchange alone,
namely due to  spin disorder 
scattering.\cite{Furukawa94,Kubo72,Kasuya56,Fisher68,Nagaev98}
It is noteworthy that the local Jahn-Teller distortion is
reported even in the metallic phase of (La,Sr)MnO$_3$,\cite{Louca97}
which might indicate that the effect of the lattice distortion 
(small polaron formation) is not so prominent for these compounds.

It has been discussed that the roles
 of other interactions such as lattice distortion\cite{Millis95}
or orbital fluctuation\cite{Ishihara96} are important in
(La,Sr)MnO$_3$. Especially, the discrepancies in
Curie temperature as well as resistivity between the double-exchange
alone model and CMR manganites are emphasized. 
However, as discussed above, accurate treatment on theoretical
calculation and the bulk measurement of single crystal samples
without grain effects show that
the double-exchange model explains various thermodynamical
properties of the high $T_{\rm c}$ compounds such as
(La,Sr)MnO$_3$.

Let us now discuss the properties of
compounds with lower $T_{\rm c}$, {\em e.g.} (La,Ca)MnO$_3$,
which qualitatively show different behaviors.
One of the major difference is that 
the resistivity above $T_{\rm c}$ shows semiconductive temperature dependence.
Several theories based on microscopic models are proposed.
Polaron effects of Jahn-Teller distortions
are introduced to explain the metal-insulator transition
from the point of view of large polaron to small polaron
crossover by magnetism.\cite{Millis96l,Roder96} 
The idea of Anderson localization due to spin 
disorder as well as diagonal charge 
disorder has also 
been discussed.\cite{Nagaev98,Varma96,MullerHartmann96,Allub96,Li97,Sheng97}

Many of these proposals relate magnetism and transport
through the change of the hopping matrix element $t_{\rm eff}(T,H)$ as 
discussed by Anderson and Hasegawa.\cite{Anderson55}
Namely, in general, these scenarios discuss the existence of the
critical value of hopping $t_{\rm c}$ determined by polaron size or
mobility edge, and explain the metallic state as large hopping
 region $t_{\rm eff}(T,H) > t_{\rm c}$ and
the insulating state as small hopping region $t_{\rm eff}(T,H) < t_{\rm c}$.
However, in order to explain the experiments for resistivity,
these scenario requires fine tuning of the parameters
in the following sense;
to explain the fact that
the metal-insulator transition always occurs in the vicinity of $T_{\rm c}$
for various A-site substitution in both ionic radius and average valence
changes, they require the pinning of the critical hopping
 $t_{\rm c} \sim t_{\rm eff}(T=T_{\rm c})$
 for any values
of carrier concentration as well as bandwidth and A-site randomness.
Note that $ t_{\rm eff}$ is determined by the short range spin correlation
$\braket{S_i\cdot S_j}$ and is a smooth and continuous function of
temperature without an anomaly at $T_{\rm c}$.
It is also required to explain the
absence of the metal-insulator transition in high-$T_{\rm c}$
compounds such as (La,Sr)MnO$_3$ at $x\sim1/3$.

In order to investigate such controversial issues,
it is important to discuss the difference of the resistivity
properties in these compounds in comparison with other 
properties from the point of view of deviation from DE behavior.
Concerning the charge dynamics,
the effect of the A-site substitution is investigated as follows.
Optical conductivity measurements\cite{Moritomo97,Jung98,Quijada98x}
show that the $\JH$-split peak at around 3eV investigated
in this paper remains as bandwidth
is changed from wide to narrow. The bandwidth control causes
changes in the spectrum at the peak structure around $\sim 1{\rm eV}$
and the infrared quasi-Drude (incoherent) structures.
In a narrow band system, Machida {\em et al.}\cite{Machida97x,Machida98x}
 has found a characteristic absorption band at $\sim$1.5 eV 
originated in transition 
of small polarons. The small polaron band disappear
 in the ferromagnetic metallic state. 

For the spin dynamics, spin wave dispersion as well as
its linewidth has been investigated.
For wide-bandwidth compounds, 
spin wave dispersion\cite{Perring96} in La$_{0.7}$Pb$_{0.3}$MnO$_3$ 
which roughly shows a cosine-band type dispersion
is explained by the DE Hamiltonian.\cite{Furukawa96,Kaplan97,Wang98}
For (La,Sr)MnO$_3$, the residual linewidth $\Gamma_0(q)$ behaves as
$\sim q^4$ in the insulating region at $x\sim 0.15$,\cite{Vasiliu-Doloc97}
which is understood as magnon-magnon scattering effect,
and as $\sim q^2$ in the metallic region at $x\sim 0.2$,\cite{Moudden98}
which is speculated to be due to inhomogeneity effect through
spin stiffness distribution.
For lower $T_{\rm c}$ compounds, there exists a systematic behavior that
broadening of the spin wave dispersion is prominent 
at the zone boundary.\cite{Hwang98}
Another unconventional feature in these compounds is the
presence of the central peak well below $T_{\rm c}$,\cite{Lynn96}
which indicates the presence of the magnetic cluster and its
diffusive dynamics. From the spin diffusion constant,
the correlation length of the spin clusters are estimated to be
$\xi \sim 10{\rm \AA}$. Neutron elastic scattering
 measurements also observed the
ferromagnetic cluster with correlation length 
$\xi\sim 20{\rm \AA}$.\cite{DeTeresa96,Fernandez-Baca98}

Thus from the points of view of charge and spin dynamics,
the observed systematic change as $T_{\rm c}$ become lower is 
considered to be the formation of magnetic cluster with
typical length scale of $\xi < 10^2 {\rm \AA}$, accompanied
by lattice distortion, or magnetoelastic polaron.
Inhomogeneities in charge and magnetic
structures\cite{Heffner96,Yoon98,Booth98}
suggest that such polaronic cluster remain even at at low temperatures.
Possible micrograin
formation due to charge segregation as well as
phase separation between ferromagnetic and antiferromagnetic domains
has also been discussed.\cite{Allodi97,Perring97,Hennion98x}

In the presence of ferromagnetic clusters, or magnetically
isolated half metallic nanodomains,
one can discuss the tunneling MR mechanism for
half metals as the origin of the CMR behavior,
as is commonly the case for polycrystals.\cite{Hwang96}
Instability of electronic phase separation\cite{Yunoki98}
is a candidate for the initial driving force for such phenomena,
which is stabilized to form droplet structures due to long range
Coulomb interactions.
Another possibility is the effect of static potential disorder
due to A-site cation $R^{3+}$-$A^{2+}$ distributions which
causes charge inhomogeneities, as well as self-trapping effect
of lattice polarons.
The idea is consistent with the phenomenological
explanation of resistivity in (La,Ca)MnO$_3$ 
by the Two-fluid model\cite{Jaime98x} which
discusses the coexistence of a metallic conductivity path and
an activation-type polaronic conductivity.

To summarize,
we investigated spin and charge excitation of DE
by both theoretical and experimental approaches.
We see the scaling behavior in spin wave lifetime
as well as interband optical spectrum.
These features are well understood by
the magnetization dependence of the DOS with half-metallic
behaviors.
Origin of the magnetoresistance is discussed as 
a spin-disorder scattering in the wide bandwidth region
and tunneling MR due to half-metallic behavior
in the narrow bandwidth region.

\section*{Acknowledgments}

N.F. is supported by the Mombusho Grant for overseas research.
N.F. thanks E. Dagotto, A. Moreo and C.M. Varma for fruitful discussions.
Y.M would like to thank A. Machida for his help in optical measurements.
The work was supported by a Grant-In-Aid from Precursory Research for Embryonic Scienece and Technology (PRESTO), Japan Science and 
Technology Cooperation (JST), Japan.

\pagebreak

\begin{thebibliography}{10}
\bibitem{Nagaev97}
Phase separation in the weak exchange limit ($\JH \ll W$)
is also discussed in E.~L. Nagaev, Physica B {\bf 230-232},  816  (1997).
Note that the ground state is not half metal in this regime.
\end{thebibliography}
\end{document}